\documentclass[pra,notitlepage]{revtex4-1}

\usepackage{amsmath}
\usepackage{times}

\begin{document}

\title{Tan Relations in Dilute Bose Gasses}

\author{Adriaan M. J.~Schakel}

\affiliation{
Laborat\'orio de F\'isica Te\'orica e Computacional,
Departamento de F\'{i}sica,
Universidade Federal de Pernambuco, 
50670-901, Recife-PE, Brazil}

\begin{abstract}
The exact relations for strongly interacting Fermi gasses, recently
derived by Tan, are shown to first order in the loop expansion to also
apply to dilute Bose gasses.  A simple thermodynamic argument is put
forward to support their validity.  As an application, the second-order
correction to the depletion of the condensate is determined to
logarithmic accuracy.
\end{abstract}

\maketitle

In a sequence of profound original papers
\cite{Tan:2008_a,Tan:2008_b,Tan:2008_c}, Tan recently derived various
exact relations for strongly correlated Fermi gasses.  Generalizing the
standard formula for a free Fermi gas, he expressed the ground-state
energy of a (stongly) interacting Fermi gas with two spin states as a
simple linear functional of the particle momentum distribution $n(k)$,
which he argued to fall off like $C/k^4$ at large wave number $k$.
Finiteness of the energy in the ultraviolet requires that this tail be
subtracted.  Tan demonstrated that the ground-state energy explicitly
depends on the coefficient $C$ of this tail.  He, moreover, showed that
this coefficient, the ``contact'', determines the change in ground-state
energy due to a small variation in the s-wave scattering length, which
he dubbed ``the adiabatic sweep theorem'', and that it features in a
relation connecting the pressure to the ground-state energy as well as
in a generalized virial theorem.  In essence, these results imply that
the sole parameter $C$ captures the correlations of a Fermi gas (with
short-range interactions).  It is remarkable that this parameter,
characterizing the behavior of interacting fermions at \emph{short}
distances ($k \to \infty$), features in thermodynamic quantities that
characterize the system at \emph{long} distances.  Two of Tan's
predictions, viz.\ the adiabatic sweep and the generalized virial
theorems, were very recently verified in experiments at JILA on ultra
cold gases of fermionic $^{40}$K atoms confined in a harmonic trapping
potential \cite{Stewert:2010}.

It was pointed out by Combescot \textit{et al.} \cite{Combescot:2009}
that the statistics of the particles plays no role in the derivation of
the Tan relations, so that they should also apply to Bose gasses.  In
this note, we explicitly show this to be the case to first order in the
loop expansion.  The exactness of the relations is supported by a simple
thermodynamic argument we put forward.  Using these relations, we
determine, for the first time, the second-order correction to the
depletion of the condensate to logarithmic accuracy.

Adapted for a dilute Bose gas of one species of particles with number
density $n$, the Tan energy relation \cite{Tan:2008_a} becomes
\cite{Combescot:2009}
\begin{equation} 
\label{tan}
\mathcal{E}_\mathrm{T}(n) = \int \frac{\mathrm{d}^3k}{(2 \pi)^3}
\epsilon(k) \left[ n(k) - \frac{C}{k^4} \right] +
\frac{\hbar^2 C}{8 \pi m a} .
\end{equation} 
Here, $\epsilon(k)= \hbar^2 k^2/2m$ is the spectrum of
\emph{noninteracting} bosons and $a$ is the s-wave scattering length
characterizing the short-range, but not necessarily weak interactions.
The additional factor of $\frac{1}{2}$ in the last term in comparison to
the corresponding term for an interacting Fermi gas with two spin states
arises because only one species of particles is present here.

Equation~(\ref{tan}) is to be compared to the standard perturbative
expression for the ground-state energy density at the absolute zero of
temperature \cite{Fetter&Walecka},
\begin{equation}  
\label{En}
\mathcal{E}(n) = \frac{2 \pi \hbar^2 a n^2}{m} + \frac{1}{2} \int
\frac{\mathrm{d}^3k}{(2 \pi)^3} \left[E(k) - \epsilon(k) - g n_0 +
  \frac{g^2 n_0^2}{2 \epsilon(k)} \right] ,
\end{equation}
obtained to first order in the loop expansion.  Instead of the free
particle spectrum, this expression dominantly features the Bogoliubov
spectrum \cite{Bogoliubov:1947} $E(k) = \sqrt{\epsilon^2(k) + 2 g n_0
  \epsilon (k)}$, where $g$ is the bare coupling constant of the local
interaction term in the Hamiltonian density
\begin{equation} 
\label{Hint}
\mathcal{H}_\mathrm{int} = \frac{1}{2} g \left(\psi^*\psi \right)^2 ,
\end{equation} 
with $\psi$ the quantum field describing the atoms.  Moreover, $n_0$
denotes the number density of particles in the condensate, which to
lowest order is given by $n_0=n$.  Since the integral in Eq.~(\ref{En})
represents a one-loop contribution, $n_0$ in the integrand may to this
order be replaced with $n$, so that the right side becomes a function of
$n$ alone.  The first term in the integrand is the zero-point energy of
the quantum field describing the Bogoliubov excitations.  This
contribution to the ground-state energy diverges in the ultraviolet and
must be regularized.  The natural way to do so in the context of Bose
gasses, and in line with the modern appoach to renormalization group
theory, is by introducing a large wave number cutoff $\Lambda$.  This
parameter physically denotes the scale beyond which the microscopic
model (\ref{Hint}) seizes to be valid.  The effect of the unknown
physics above the cutoff is incorporated by redefining, or renormalizing
the parameters of the original theory so that physical observables
become finite when expressed in terms of these renormalized parameters.
The remaining three so-called counter terms in the integrand of
Eq.~(\ref{En}), which render the integral finite in the limit $\Lambda
\to \infty$, serve this purpose.  The first counter term, which
subtracts the zero-point energy of a free boson field, amounts to an
irrelevant additive constant $\propto \Lambda^5$ independent of the
parameters of the theory.  (In nonrelativistic theories, the mass
parameter is just an atomic constant as far as renormalization is
concerned.)  The second and third counter terms renormalize respectively
the chemical potential, which has been swapped for the particle number
density in Eq.~(\ref{En}), and the coupling constant \cite{BBS}.
Specifically,
\begin{subequations}
\begin{eqnarray}  
\label{bec:reno}
\mu_\mathrm{r} &\equiv& \mu - \frac{1}{12\pi^2} g \Lambda^3,
\\ g_\mathrm{r} &\equiv& g - g^2 \int \frac{\mathrm{d}^3k}{(2 \pi)^3}
\frac{1}{2 \epsilon(k)} \nonumber \\ &=& g - \frac{1}{2\pi^2}
\frac{m}{\hbar^2} g^2 \Lambda ,
\end{eqnarray}  
\end{subequations}
where renormalized parameters are given the subscript ``r''.  The
renormalized coupling constant is related to the s-wave scattering
length $a$ through $g_\mathrm{r} = 4 \pi \hbar^2 a/m$.  This has been
used in the first term at the right side of Eq.~(\ref{En}), $\frac{1}{2}
g_\mathrm{r} n^2$, which represents the contribution from the
condensate.  Evaluation of the wave vector integrals in Eq.~(\ref{En}) yields
in the limit $\Lambda \to \infty$ the celebrated result due
to Lee and Yang \cite{Lee:1957}
\begin{equation} 
\label{LeeYang}
\mathcal{E}(n) = \frac{2 \pi \hbar^2 a n^2}{m} \left[1 + \frac{128}{15 }
  \left(\frac{a^3 n}{\pi}\right)^{1/2} \right] ,
\end{equation} 
where, consistent to one-loop oder, also $g$ in the one-loop integral has
been replaced with $g_\mathrm{r}$.

As an aside, we remark that in dimensional regularization no counter
terms are needed at all in Eq.~(\ref{En}), for positive powers of the
cutoff do not show up in this scheme \cite{BBS}.

The particle momentum distribution $n(k)$ featuring in Tan's
energy relation (\ref{tan}) reads to this order \cite{Fetter&Walecka}
\begin{equation} 
\label{dis}
n(k) = \frac{1}{2} \left(\frac{\epsilon(k) + g
  n_0}{E(k)} -1 \right) .
\end{equation} 
It varies at large wave numbers in accordance with Tan's observation as
$n(k) \sim C_0/k^4$, with $C_0 \equiv \left(m g n_0/\hbar^2\right)^2$.
Upon substituting this for $C$ in the integrand of Eq.~(\ref{tan}),
which is justified to one-loop order, and carrying out the wave vector
integrals, we obtain in the limit $\Lambda \to \infty$
\begin{equation} 
\label{ET}
\mathcal{E}_\mathrm{T}(n) = -\frac{128 \pi^{1/2}}{5} \frac{\hbar^2 
  (a n)^{5/2}}{m} + \frac{\hbar^2 C}{8 \pi m a} .
\end{equation} 
To demonstrate that this agrees with the Lee-Yang result
(\ref{LeeYang}), we compute the contact $C$ to first order in the loop
expansion.

To this end, we use the result derived by Braaten and Platter
\cite{Braaten:2008} through the operator product expansion that the
contact is determined by the expectation value of the interaction term
(\ref{Hint})
\begin{equation}
\label{OPE} 
C = \left(\frac{m g}{\hbar^2}\right)^2 \left\langle \left(\psi^*\psi
\right)^2 \right\rangle .
\end{equation} 
This result was originally derived for a Fermi gas, but since the
derivation is independent of the statistics of the particles, it also
applies here.  To account for the condensate, the (complex) field $\psi$
is shifted as follows
\begin{equation} 
\psi = v + \frac{1}{\sqrt{2}} \left(\chi_1 +
  \mathrm{i} \chi_2\right),
\end{equation} 
with $\chi_1(t,\mathbf{x})$ and $\chi_2(t,\mathbf{x})$ two real fields.
The expectation value $v$ of the field $\psi$ is related to the number
density $n_0$ of condensed particles through $v^2 = n_0$.  Without loss
of generality, we assumed that $v$ is real.  The propagator of the
multiplet $(\chi_1, \chi_2)^\mathrm{T}$ as a function of frequency
$\omega$ and wave vector $\mathbf{k}$ is given at lowest order by
\begin{equation} 
\label{bec:prop}
\Delta_\mathrm{F}(\omega, \mathbf{k}) = \frac{1}{\hbar^2 \omega^2 -
  E^2({\bf k}) + \mathrm{i} \eta } \left(
\begin{array}{cc} 
 \epsilon({\bf k}) & \mathrm{i} \hbar \omega \\ - \mathrm{i} \hbar \omega&
 \epsilon({\bf k}) + 2 g n_0 
\end{array} \right)  ,                                          
\end{equation} 
with $\eta>0$ the usual infinitesimal parameter, introduced for
causality, that is to be set to zero at the end of the calculation, and
$E({\bf k})$ the Bogoliubov spectrum.  To the one-loop order,
$\left\langle \left(\psi^*\psi \right)^2 \right\rangle = n_0^2 + 3 n_0
\left\langle \chi_1^2 \right\rangle + n_0 \left\langle \chi_2^2
\right\rangle$ so that by Eq.~(\ref{bec:prop})
\begin{equation} 
C = \left(\frac{m g}{\hbar^2}\right)^2 n_0 \left(n_0 + \frac{3}{2} \int
\frac{\mathrm{d}^3k}{(2 \pi)^3} \frac{\epsilon(k)}{E(k)} + \frac{1}{2}
\int \frac{\mathrm{d}^3k}{(2 \pi)^3} \frac{E(k)}{\epsilon(k)} \right)
\end{equation} 
after carrying out the frequency integrals.  Note that the wave vector
integrals appearing here diverge in the ultraviolet.  Apart from an
irrelevant diverging term $\propto \Lambda^3$ independent of the
parameters of the theory, the divergences can be canceled as in the
fermionic theory \cite{Braaten:2008} by going over to the renormalized
coupling constant, giving
\begin{equation} 
\label{vev} 
C = \left(\frac{m g_\mathrm{r} n_0}{\hbar^2}\right)^2 \left[1+
  \frac{80}{3} \left(\frac{a^3 n}{\pi}\right)^{1/2} \right].
\end{equation} 
The density of particles $n_0$ residing in the condensate is given to
this order by the Bogoliubov depletion formula \cite{Bogoliubov:1947}
\begin{eqnarray} 
n_0 &=& n - \int \frac{\mathrm{d}^3k}{(2 \pi)^3} n(k) \nonumber \\
&=& n \left[1 - \frac{8}{3} \left(\frac{a^3 n}{\pi}\right)^{1/2} \right] .
\end{eqnarray} 
Use of this in Eq.~(\ref{vev}) yields for the contact to first order in
the loop expansion
\begin{equation}
\label{C} 
C = (4 \pi a n)^2 \left[1+ \frac{64}{3} \left(\frac{a^3 n}{\pi}\right)^{1/2}
  \right] .
\end{equation} 
Inserting this expression for $C$ into Eq.~(\ref{ET}), we recover the
Lee-Yang result (\ref{LeeYang}) and thereby showed that, to first order
in the loop expansion, the Tan energy relation (\ref{tan}) agrees with
the Bogoliubov theory.

Given the result (\ref{C}) for the contact, it is readily verified that
Tan's adiabatic and pressure relations
\cite{Tan:2008_b},
\begin{equation} 
\label{adiabatic}
\frac{\mathrm{d} \mathcal{E}}{\mathrm{d} a} = \frac{\hbar^2 C}{8 \pi m
  a^2}, \quad \quad P = \frac{2}{3} \mathcal{E} + \frac{\hbar^2 C}{24 \pi m a},
\end{equation} 
are satisfied by the standard expressions obtained from Bogoliubov
theory.  The last terms in these expressions carry again an additional
factor of $\frac{1}{2}$ in comparison to their fermionic counterparts
because only one species of particles is present here.  The two
(nonperturbative) relations (\ref{adiabatic}) can be combined to yield
for the pressure
\begin{equation}
\label{P1} 
P = \frac{1}{3a} \frac{\mathrm{d}}{\mathrm{d} a} \left( a^2 \mathcal{E}
\right) .
\end{equation} 
This expression is closely related to the thermodynamic relation
\begin{equation} 
\label{P2}
P = n^2 \frac{\mathrm{d}}{\mathrm{d} n} \left( \frac{\mathcal{E}}{n} \right) 
\end{equation} 
and can in fact be derived from it by noting that the leading term in
the ground-state energy is proportional to $a n^2$, while the quantum
corrections are, on dimensional grounds, a function of $a^3 n$ alone.
This simple, yet nonperturbative argument gives strong support to the
validity of the Tan relations (\ref{adiabatic}).

We next extend the analysis to second order in the loop expansion.  The
second-order expression for the ground-state energy
\cite{Wu:1959,Hugenholtz:1959,Sawada:1959} yields, to logarithmic
accuracy, the extra term $(32/3) \left( 4 \pi - 3 \sqrt{3} \right) a^3 n
\ln(a^3 n)$ within the square brackets in Eq.~(\ref{C}). The calculation
of the expectation value in Eq.~(\ref{OPE}) involves, apart from minor
changes in the vertices, the same diagrams that contribute to the
thermodynamic potential, which have been treated in detail in
Ref.~\cite{Braaten:1999}.  To logarithmic accuracy, we find
\begin{equation} 
\label{vev2} 
\left\langle \left(\psi^*\psi \right)^2 \right\rangle = n_0^2 \left[1 +
  \frac{80}{3} \left(\frac{ a^3 n}{\pi}\right)^{1/2} + \frac{16}{3}
  \left( 4 \pi - 3 \sqrt{3} \right) a^3 n \ln( a^3 n) \right] .
\end{equation} 
Inserting these results into Eq.~(\ref{OPE}), we obtain as next-order
correction to the depletion of the condensate 
\begin{equation} 
n_0 = n \left[1 - \frac{8}{3} \left(\frac{a^3 n}{\pi}\right)^{1/2} +
  \frac{8}{3} \left( 4 \pi - 3 \sqrt{3} \right) a^3 n \ln(a^3 n)
  \right] .
\end{equation} 
A direct calculation of this correction term, without using Tan
relations, appears laborious and has, to our knowledge, not been
reported in the literature.

At the face of it, Tan's energy relation (\ref{tan}) is surprising from
the theory side in that it makes no explicit reference to the
condensate.  This is in sharp contrast to the standard perturbative
approach where the condensate is explicitly accounted for from the
onset, see Eq.~(\ref{En}).  Given the experimental success in verifying
the Tan relations in strongly interacting Fermi gasses, a similar
experimental undertaking in the context of dilute Bose gasses seems
desirable to gain further insights into strongly interacting systems.

\begin{acknowledgments}
The author is indebted to G.~L.~Vasconcelos for warm hospitality at the
Departamento de F\'{i}sica, Universidade Federal de Pernambuco, Recife.
Financial support from CAPES, Brazil through a visiting professor
scholarship is gratefully acknowledged.
\end{acknowledgments}

\bibliographystyle{apsrev}
\bibliography{adb}

\begin{thebibliography}{14}
\expandafter\ifx\csname natexlab\endcsname\relax\def\natexlab#1{#1}\fi
\expandafter\ifx\csname bibnamefont\endcsname\relax
  \def\bibnamefont#1{#1}\fi
\expandafter\ifx\csname bibfnamefont\endcsname\relax
  \def\bibfnamefont#1{#1}\fi
\expandafter\ifx\csname url\endcsname\relax
  \def\url#1{\texttt{#1}}\fi
\expandafter\ifx\csname urlprefix\endcsname\relax\def\urlprefix{URL }\fi
\providecommand*{\bibinfo}[2]{#2}
\providecommand*{\eprint}[1]{#1}
\providecommand*{\url}[1]{#1}
\begingroup\makeatletter
 \@temptokena{%
  \expandafter\ifx\csname citenamefont\endcsname\relax
   \DeclareRobustCommand\citenamefont{\@firstofone}%
   \global\let\citenamefont\citenamefont
   \global\expandafter\let\csname citenamefont \expandafter\endcsname\csname
  citenamefont \endcsname
  \fi
 }\if@filesw\immediate\write\@auxout{\the\@temptokena}\fi
\expandafter\endgroup\the\@temptokena

\bibitem[{\citenamefont{Tan}(2008{\natexlab{a}})}]{Tan:2008_a}
\bibinfo{author}{\bibfnamefont{S.}~\bibnamefont{Tan}}, \bibinfo{journal}{Ann.
  Phys. (N.Y.)} \textbf{\bibinfo{volume}{323}}, \bibinfo{pages}{2952}
  (\bibinfo{year}{2008}{\natexlab{a}}).

\bibitem[{\citenamefont{Tan}(2008{\natexlab{b}})}]{Tan:2008_b}
\bibinfo{author}{\bibfnamefont{S.}~\bibnamefont{Tan}}, \bibinfo{journal}{Ann.
  Phys. (N.Y.)} \textbf{\bibinfo{volume}{323}}, \bibinfo{pages}{2971}
  (\bibinfo{year}{2008}{\natexlab{b}}).

\bibitem[{\citenamefont{Tan}(2008{\natexlab{c}})}]{Tan:2008_c}
\bibinfo{author}{\bibfnamefont{S.}~\bibnamefont{Tan}}, \bibinfo{journal}{Ann.
  Phys. (N.Y.)} \textbf{\bibinfo{volume}{323}}, \bibinfo{pages}{2987}
  (\bibinfo{year}{2008}{\natexlab{c}}).

\bibitem[{\citenamefont{Stewart} \emph{et~al.}(2010)\citenamefont{Stewart,
  Gaebler, Drake, and Jin}}]{Stewert:2010}
\bibinfo{author}{\bibfnamefont{J.~T.} \bibnamefont{Stewart}},
  \bibinfo{author}{\bibfnamefont{J.~P.} \bibnamefont{Gaebler}},
  \bibinfo{author}{\bibfnamefont{T.~E.} \bibnamefont{Drake}}, \bibnamefont{and}
  \bibinfo{author}{\bibfnamefont{D.~S.} \bibnamefont{Jin}},
  \bibinfo{journal}{Phys. Rev. Lett.} \textbf{\bibinfo{volume}{104}},
  \bibinfo{pages}{235301} (\bibinfo{year}{2010}).

\bibitem[{\citenamefont{Combescot} \emph{et~al.}(2009)\citenamefont{Combescot,
  Alzetto, and Leyronas}}]{Combescot:2009}
\bibinfo{author}{\bibfnamefont{R.}~\bibnamefont{Combescot}},
  \bibinfo{author}{\bibfnamefont{F.}~\bibnamefont{Alzetto}}, \bibnamefont{and}
  \bibinfo{author}{\bibfnamefont{X.}~\bibnamefont{Leyronas}},
  \bibinfo{journal}{Phys. Rev.} \textbf{\bibinfo{volume}{A79}},
  \bibinfo{pages}{053640} (\bibinfo{year}{2009}).

\bibitem[{\citenamefont{Fetter and Walecka}(1971)}]{Fetter&Walecka}  See, for example,
\bibinfo{author}{\bibfnamefont{A.~L.} \bibnamefont{Fetter}} \bibnamefont{and}
  \bibinfo{author}{\bibfnamefont{J.~D.} \bibnamefont{Walecka}},
  \emph{\bibinfo{title}{Quantum Theory of Many-Particle Systems}}
  (\bibinfo{publisher}{McGraw-Hill}, \bibinfo{address}{New York, N.Y.},
  \bibinfo{year}{1971}).

\bibitem[{\citenamefont{Bogoliubov}(1947)}]{Bogoliubov:1947}
\bibinfo{author}{\bibfnamefont{N.~N.} \bibnamefont{Bogoliubov}},
  \bibinfo{journal}{J. Phys. (Moscow)} \textbf{\bibinfo{volume}{11}},
  \bibinfo{pages}{23} (\bibinfo{year}{1947}).

\bibitem[{\citenamefont{Schakel}(2008)}]{BBS}  For details, see the textbook:
\bibinfo{author}{\bibfnamefont{A.~M.~J.} \bibnamefont{Schakel}},
  \emph{\bibinfo{title}{Boulevard of Broken Symmetries: Effective Field
  Theories of Condensed Matter}} (\bibinfo{publisher}{World Scientific},
  \bibinfo{address}{Singapore}, \bibinfo{year}{2008}).

\bibitem[{\citenamefont{Lee and Yang}(1957)}]{Lee:1957}
\bibinfo{author}{\bibfnamefont{T.~D.} \bibnamefont{Lee}} \bibnamefont{and}
  \bibinfo{author}{\bibfnamefont{C.~N.} \bibnamefont{Yang}},
  \bibinfo{journal}{Phys. Rev.} \textbf{\bibinfo{volume}{105}},
  \bibinfo{pages}{1119} (\bibinfo{year}{1957}).

\bibitem[{\citenamefont{Braaten and Platter}(2008)}]{Braaten:2008}
\bibinfo{author}{\bibfnamefont{E.}~\bibnamefont{Braaten}} \bibnamefont{and}
  \bibinfo{author}{\bibfnamefont{L.}~\bibnamefont{Platter}},
  \bibinfo{journal}{Phys. Rev. Lett.} \textbf{\bibinfo{volume}{100}},
  \bibinfo{pages}{205301} (\bibinfo{year}{2008}).

\bibitem[{\citenamefont{Wu}(1959)}]{Wu:1959}
\bibinfo{author}{\bibfnamefont{T.~T.} \bibnamefont{Wu}},
  \bibinfo{journal}{Phys. Rev.} \textbf{\bibinfo{volume}{115}},
  \bibinfo{pages}{1390} (\bibinfo{year}{1959}).

\bibitem[{\citenamefont{Hugenholtz and Pines}(1959)}]{Hugenholtz:1959}
\bibinfo{author}{\bibfnamefont{N.~M.} \bibnamefont{Hugenholtz}}
  \bibnamefont{and} \bibinfo{author}{\bibfnamefont{D.}~\bibnamefont{Pines}},
  \bibinfo{journal}{Phys. Rev.} \textbf{\bibinfo{volume}{116}},
  \bibinfo{pages}{489} (\bibinfo{year}{1959}).

\bibitem[{\citenamefont{Sawada}(1959)}]{Sawada:1959}
\bibinfo{author}{\bibfnamefont{K.}~\bibnamefont{Sawada}},
  \bibinfo{journal}{Phys. Rev.} \textbf{\bibinfo{volume}{116}},
  \bibinfo{pages}{1344} (\bibinfo{year}{1959}).

\bibitem[{\citenamefont{{Braaten} and {Nieto}}(1999)}]{Braaten:1999}
\bibinfo{author}{\bibfnamefont{E.}~\bibnamefont{{Braaten}}} \bibnamefont{and}
  \bibinfo{author}{\bibfnamefont{A.}~\bibnamefont{{Nieto}}},
  \bibinfo{journal}{Eur. Phys. J.} \textbf{\bibinfo{volume}{B11}},
  \bibinfo{pages}{143} (\bibinfo{year}{1999}).

\end{thebibliography}

\end{document}